\documentclass[final,5p,times,twocolumn]{elsarticle}
\usepackage{graphicx}
\usepackage{dcolumn}
\usepackage{bm}
\usepackage{color}
\usepackage{amssymb}
\usepackage{amsthm}
\usepackage{setspace}

\newcommand{\be}{\begin{equation}}
\newcommand{\ee}{\end{equation}}
\newcommand{\ba}{\begin{eqnarray}}
\newcommand{\ea}{\end{eqnarray}}
\newcommand{\bd}{\begin{displaymath}}
\newcommand{\ed}{\end{displaymath}}
\newcommand{\bea}{\begin{eqnarray}}
\newcommand{\eea}{\end{eqnarray}}

\journal{Physics Letters A}

\begin{document}

\begin{frontmatter}

\title{Laser Wake Field Collider}

\author{
Istv\'an Papp $^{1,2}$,
Larissa Bravina $^3$,
M\'aria Csete  $^4$,
Igor N. Mishustin  $^{5,6}$, 
D\'enes Moln\'ar $^{7}$,\\
Anton Motornenko   $^5$, 
Leonid M. Satarov  $^{5}$,
Horst St\"ocker  $^{5,8,9}$,
Daniel D. Strottman $^{10}$,\\
Andr\'as Szenes  $^4$,
D\'avid Vass   $^4$,
Tam\'as S. Bir\'o $^1$,
L\'aszl\'o P. Csernai $^{1,2,5}$,
Norbert Kro\'o $^{1,11}$\\
(NAPLIFE Collaboration)\\}
\address{
$^{1}$ Wigner Research Centre for Physics, Budapest, Hungary\\
$^2$ Dept. of Physics and Technology, University of Bergen, 5007 Bergen, Norway\\
$^3$ Department of Physics, University of Oslo, Norway\\
$^4$ Dept. of Optics and Quantum Electronics, Univ. of Szeged, Hungary\\
$^5$ Frankfurt Institute for Advanced Studies, 60438 Frankfurt/Main, Germany\\
$^6$ National Research Center "Kurchatov Institute" Moscow, Russia\\
$^7$ Dept. of Physics, Purdue University, West Lafayette, 47907 IN, USA \\
$^8$ Inst. f\"ur Theoretische Physik, Goethe Universit\"at Frankfurt, 60438 Frankfurt/Main, Germany\\
$^9$ GSI Helmholtzzentrum f\"ur Schwerionenforschung GmbH, 64291 Darmstadt, Germany\\
$^{10}$ Los Alamos National Laboratory, Los Alamos, 87545 NM, USA \\
$^{11}$ Hungarian Academy of Sciences, 1051 Budapest, Hungary\\[-2em]
}

\begin{abstract}
Recently NAno-Plasmonic, Laser Inertial Fusion 
Experiments (NAPLIFE) were proposed, as an improved way to
achieve laser driven fusion. The improvement is the combination of two
basic research discoveries: (i) the possibility of detonations on space-time
hyper-surfaces with time-like normal (i.e. simultaneous detonation in a whole 
volume) and (ii) to increase this volume to the whole target, by regulating 
the laser light absorption using nanoshells or nanorods as antennas.
These principles can be realized in a one dimensional configuration,
in the simplest way with two opposing laser beams as in particle colliders.
Such, opposing laser beam experiments were also performed recently. 
  
Here we study the consequences of the Laser Wake Field Acceleration (LWFA)
if we experience it in a colliding laser beam set-up. These studies can
be applied to laser driven fusion, but also to other rapid phase transition,
combustion, or ignition studies in other materials.
\end{abstract}

\begin{keyword}
	laser wake-field acceleration, ionisation, inertial confinement fusion, NAPLIFE
	
\end{keyword}
\end{frontmatter}

\section{Introduction}

In recent years the Laser Wake Field Acceleration (LWFA) became a
well known concept with useful applications. An intensive laser pulse
impinging on a target creates a high density plasma of ~ 4 x 
10$^{19}$ / cm$^3$, and a wake field wave follows the pulse. 
This, non-linear wave in dense plasma is formed of the 
EM-field, electrons and ions. A typical laser  of 20 mJ pulse energy, 
7 fs length and $\lambda$ wavelength can create a Laser Wake Field
(LWF) dense plasma wave of about 10 $\lambda$ wavelength.

This wave is different from radio transmission waves in air or vacuum,
where the material is dilute, not or weakly ionized. Still there are
interesting phenomena if radio transmission waves create an interference.
In the 1950s this radio wave interference was well known due to 
radio jamming (e.g. jamming the Radio Free Europe shortwave AM 49m band
broadcast in Eastern Europe). Strong, unmodulated jamming broadcast 
on the same carrier frequency could lead to noiseless quiet sound, or
it was white-noise modulated resulting in strong noise. Interference 
of two original frequencies that are quite close can lead to a ``beat 
frequency transmission" with $f_{beat} = (f_1 - f_2)/2$. This is often 
too low to be perceived as an audible tone or pitch, instead, it is 
perceived as a periodic variation in the amplitude of the broadcast.
We aim to study similar kind of variety of possibilities in
colliding LWF waves.

Laser Wake Field Collider (LWFC) waves can be realized the 
simplest way by two opposing
laser light on a target. This was recently suggested in ref. 
\cite{CsEA2020,CsKP2018}
for laser driven fusion. Here two known effects were combined. First, 
the possibility of detonations on space-time
hyper-surfaces with time-like normal, so called time-like detonations
\cite{Cs1987,CS2015},
which were found theoretically and experimentally in high energy heavy
ion collision in the couple of last decades. This simultaneous volume
ignition eliminates the possibility of Rayleigh-Taylor instabilities,
which is a serious obstacle in achieving laboratory scale nuclear fusion.
The second effect we use is to achieve simultaneous detonation in the whole 
volume of the target, by regulating the laser light absorption with 
nanoshells or nanorods
\cite{KR2016,CMea2020}.

Using these two ideas a one-dimensional fusion configuration was 
suggested in ref.
\cite{CsEA2020},
with two opposing energetic laser beams.

\begin{figure*}[!t]  
\begin{center}
\resizebox{0.9\textwidth}{!}
{\includegraphics{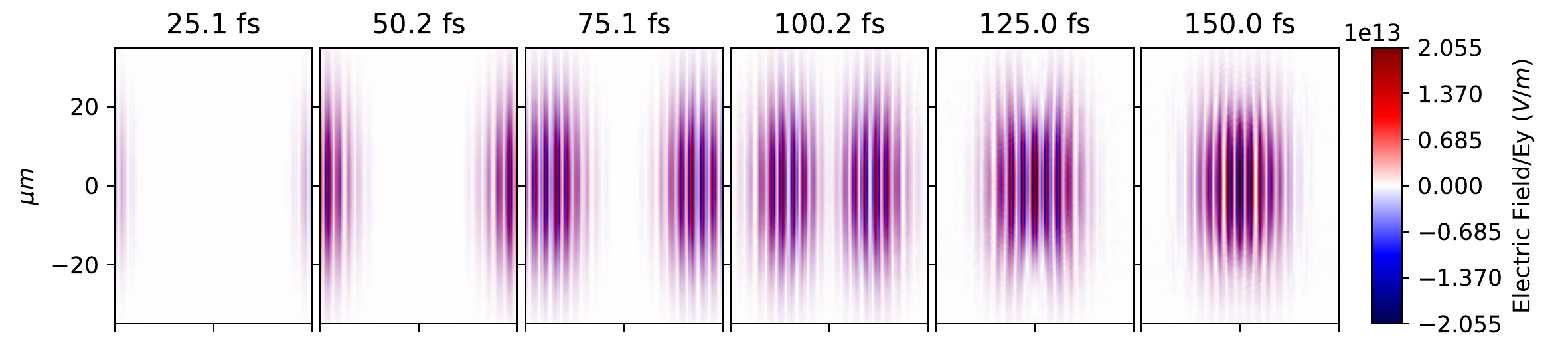}}\\
\resizebox{0.9\textwidth}{!}
{\includegraphics{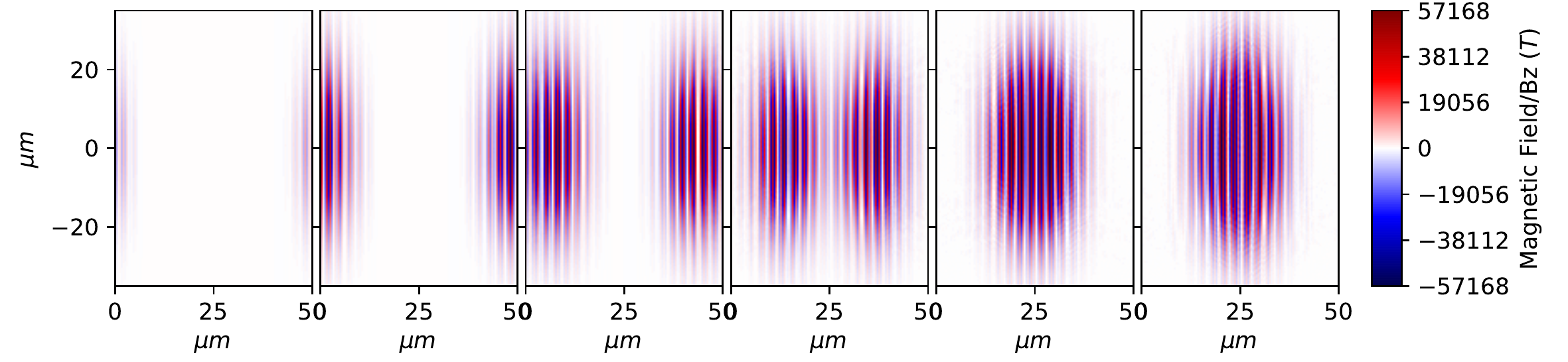}}
\end{center}
\vskip -4mm
{\caption{(color online) The electric field, $E_y$ (top) and 
magnetic field, $B_z$ (bottom) show two a Laser Wake Field (LWF)
waves approaching each other (see at 100 fs) formed by irradiation from the $\pm x$- direction.
The rest number density of the $H$ target is 
$n_H = 2.13 \cdot 10^{25} / $m$^3 = 2.13 \cdot 10^{19} / $cm$^3$.
The laser beam wavelength is $\lambda = 1 \mu$m.
The LWF wavelength is about 20 $\lambda$.}}
\label{Ex-one}
\end{figure*} 

Previous works also used colliding laser beams and LWFA for 
electron acceleration. In ref. 
\cite{Faure2006}
with a gas-jet target,
the primary stronger beam created the LWF bubble and a 
three times weaker counter-propagating beam injected 
electrons into the bubble. The ultrashort laser pulses 
had the same central wavelength and polarization.
The electron beams obtained in this way are collimated (5 mrad
divergence), mono energetic (with energy spread of 10 per cent),
and electron bunch durations shorter than 10 fs.

Similar method is presented in ref.
\cite{Hansson2016} 
for electron acceleration, also in asymmetric configuration, where the
weaker injection beam was not collinear but counter-propagating under an angle of 150$^\circ$.
These colliding laser beams did not aim for any change of the
target and have not used nanoplasmonics or simultaneous volume 
transition.

A different method for electron accelerator was presented in ref.
\cite{Bar-Lev2014}
where two identical laser beams irradiated an array of nanoantennas,
which were nanorods with long axis aligned parallel to the polarization of the laser light.
Electron bunches were hitting the layer of these nanoantennas for 
the side aligned with the directions of the nanorods. These bunches
were then accelerated each time passing a nanorod with the right
frequency and bunch period length. Although this proposal utilized 
nanoantennas, but for the local accelerating electric field to
accelerate electrons. The target was a static nanoantenna array,
changing its phase due to the incoming laser pulses.

Another experiment using colliding laser beams
\cite{Bonasera2019}
aimed for achieving high target density for nuclear fusion,
but did not utilize LWF waves and nanoplasmonics and did not
attempt to have simultaneous transition in the target.

Thus, our present aim differs from these previous 
works.

\section{Non-thermal Ignition Rate}
\label{NIR}

In Laser Wake Field Collider, the target has two sides, which are
initially accelerated towards each other. 

Carbon C6+ ions in the 
relativistic-induced transparency acceleration (RITA) regime,
have reached near to 1 GeV energy on 300 nm target thickness
\cite{Jung2015}.

In ref.
\cite{FLZhang2011}
it was shown that
intense laser pulse irradiating a combination target
can accelerate carbon ions to the TeV level by the 
laser plasma wakefield.

If we consider LWFC with a double layer 
target\footnote{The initial target 
may be two layers, e.g. D and T, with a gap between.}
pre-compressed to ion density,  $n_{pc}$, and pre-accelerated 
to several GeV/nucleon energy, (i.e. to a Laser Wake (LW) 
velocity near to the speed of light, $v_{LW} \approx\ c$),
the two LWF waves can inter-penetrate and lead to an ignition
reaction rate of
\be
     2\ \gamma^2\ n_{pc}^2\ c\ \sigma \, ,
\ee
where $\sigma$ is the ion-ion cross section, and due to the Lorentz 
contraction, the two ion bunches are compressed to  
$\gamma\ n_{pc}$. This  may well exceed the thermal (th) rate per particle
$n \langle v_{th} \sigma \rangle $.
If the ions are accelerated to 5 GeV/nucleon, then $\gamma \approx 6$,
and if the pre-compression reaches a factor 8 (considerably less
than at NIF, where the 3-dimensional compression reaches 800 g/cm$^3$), then
our burning rate is 
$
2 \ \gamma^2\ \frac{\, 8^2 n^2}{800\,n}\ (c/v_{th}) \approx 270
$
times bigger than at NIF. Here we assumed that the average thermal 
collision speed is $v_{th} \approx c/2$.
So, the non-thermal non-equilibrium Laser Wake Field Collider mechanism
may well exceed the thermal ignition rate by the adiabatic compression
and heating at NIF. Especially if this ignition takes place for a
central hot-spot only and the flame has to propagate over the
rest of the target.

The other advantage of the LWFC configuration that we can have exclusively
the most optimal DT reactions, without DD and TT ones, like in a 
mixed DT target.

\begin{figure*}[t!]  
\begin{center}
\resizebox{0.9\textwidth}{!}
{\includegraphics{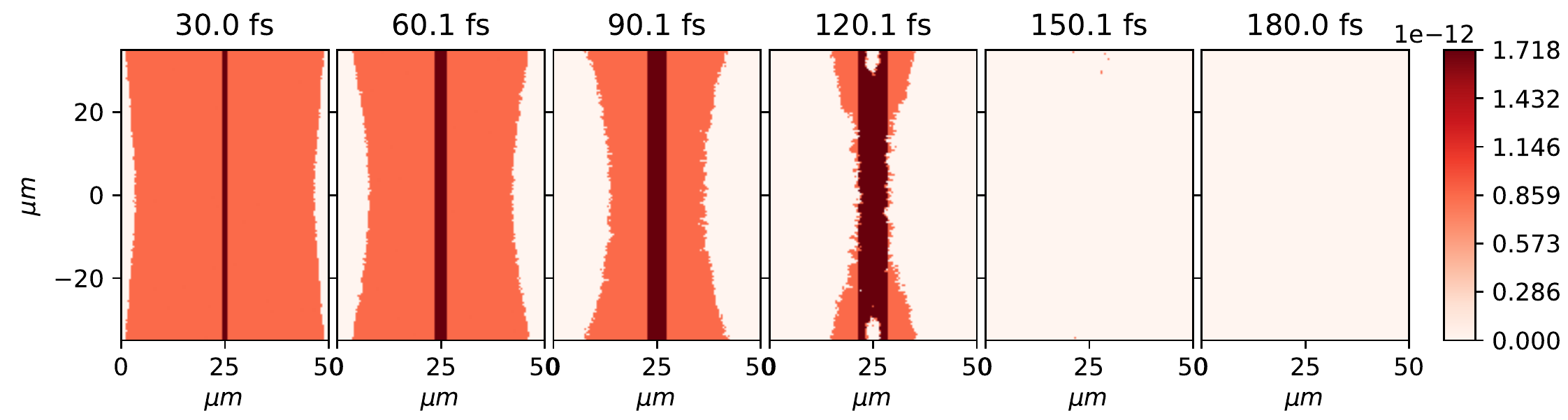}}
\end{center}
\vskip -5mm
\caption{(color online) The ionization of the $H$ atoms 
in a Laser Wake Field (LWF) wave due to the irradiation from both 
the $\pm x$- directions, on an initial target density of
$n_H = 2.13 \cdot 10^{27} $  atoms$/ $m$^3 = 2.13 \cdot 10^{21} $  atoms$/ $cm$^3$.
The energy of the $H$ atoms in Joule [J] per marker particle is shown.
The $H$ atoms disappear as protons and electrons are created.
Due to the initial momentum of the colliding $H$ slabs, the target and
projectile slabs interpenetrate each other and this leads to double 
energy density.
Several time-steps are shown at 30 fs time difference.}
\label{Ex-both}
\end{figure*}

\section{Laser and Target parameters}
\label{LTP}

Irradiating a dense target with such beams creates LWF waves, and
we study what are the consequences. We used the EPOCH multi-component
PIC code \cite{Arber2015} to see first what kind of  LWF waves develop, if we 
irradiate a target with a laser beam having a
wavelength of $\lambda = 1 \mu$m,\
a Gaussian distribution in the transverse, $[y,z]$ plane of 
half amplitude time $\delta_{t} = 26$fs,
and full pulse length $\Delta_{t} = 52$fs.

The laser focus average diameter is $2R = 40 \mu$m.
The laser pulse energy in the present test 
is 19.6 J, and the maximum intensity, in the
center of the transverse plane at the top intensity time is
$3.0 \cdot 10^{19}$ W/cm$^2$.

We studied the EM field
and the target in a 3-dimensional box of 
50 x 70 x 70 $\mu$m, divided into cubic cells of 100 nm (0.1 $\mu$m).
The initial target consisted of $H$ atoms. The $H$ target consisted of 
two slabs with 24.5 $\mu$m thickness with a 1 $\mu$m gap between them,
and pre-accelerated towards each other with a momentum of 100 MeV/c. 
Due to the
irradiation the target became mostly ionized, and protons $p$ and 
electrons, $e$ are formed.
These components were represented by marker particles. 
The rest number density of $H$ atoms in marker particles in this
work is
 $n_H = 2.13 \cdot 10^{27} $  atoms$/ $m$^3 = 2.13 \cdot 10^{21} $  atoms$/ $cm$^3$, or
 $n_H = 2.13 \cdot 10^{25} $  atoms$/ $m$^3 = 2.13 \cdot 10^{19} $  atoms$/ $cm$^3$.
This is near to the density of liquid hydrogen, which is
$2.124 \cdot 10^{22} / $cm$^3$.
The initial size of these marker particles is 
0.025$^3\ \cdot \ \mu$m$^3 = 1.5625  \cdot 10^{-5} \mu$m$^3$.
For comparison we also made a test with a two orders of magnitude
more dilute target. 
The different
types of marker particles (mp) contained different number of components:\\
$$
1 e_{mp} \propto 10^{24} e, \ \ \
1 p_{mp} \propto 10^{24} p, \ \ \
1 H_{mp} \propto 10^{24} H, \ \ \
$$
Initially the $H$ component marker particles were uniformly 
distributed in the calculation cells, so that each cell contained
$6 H_{mp}$ .

\section{Laser Wake Field Waves}
\label{LWFW}

As we can see in Fig. \ref{Ex-one} at 100.2 fs the LWF wave length is
$\lambda_{LWF} \approx 20 \mu$m. We can also see the {\bf pinch effect}, that
the transverse extent of the beam shrinks. At 150 fs the transverse 
size of the beam is about 80\% of the size at 50 fs.

\begin{figure*}[!h]  
\begin{center}
\resizebox{0.9\textwidth}{!}
{\includegraphics{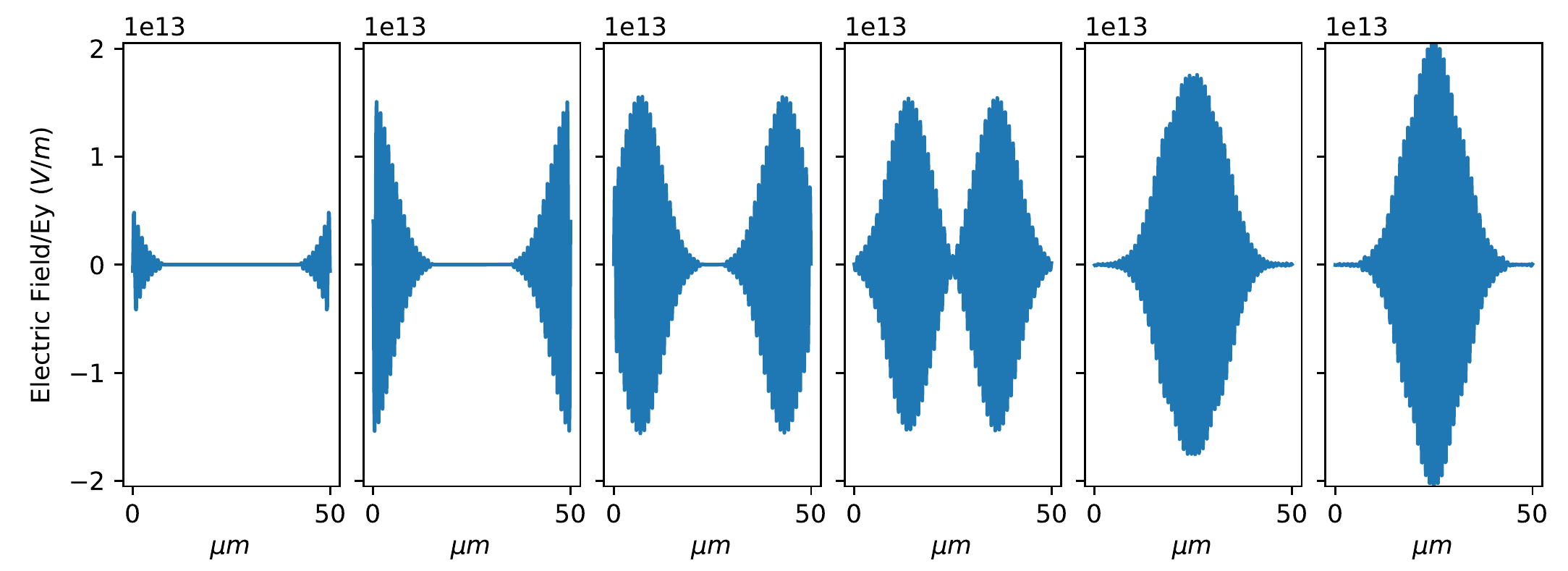}}
\end{center}
\caption{(color online) Electric Field, E$_y$ 
in a Laser Wake Field (LWF)
wave formed by irradiation from both the $\pm x$- directions. 
The field strengths are shown in the middle of the transverse, $y,z$, 
plane along the $x$-axis.
In the other, not shown, directions the fields are 
weaker by orders of magnitude.
The initial target density is $n_H = 2.13 \cdot 10^{19} $  atoms$/ $cm$^3$.
Several time-steps are shown at 25, 50, 75, 100, 125 and 150 fs times.}
\label{A-test-1}
\end{figure*}

In Fig. \ref{Ex-both} we see the effect of two opposing laser beams on the 
two target $H$ atom slabs. The two beams hit the denser targets from opposite
sides, and as the irradiation is absorbed the $H$ atoms are ionized, protons and electrons become free. By 150 fs, the $H$ atom targets are fully ionized,
except at the outside edges where the pinched irradiation beams did not
hit the target.

Electrons were created during the irradiation. $H$ ions and electrons
were moving during the calculation and crossed over to other Cells.

In Fig. \ref{A-test-1}
we can see that at around 125 - 150 fs at this lower target density,
the two LWF waves constructively
interact and the EM field strength is maximal. This moment of time would 
be adequate for a short, intensive ignition pulse.

\begin{figure*}[!t]  
\begin{center}
\resizebox{0.9\textwidth}{!}
{\includegraphics{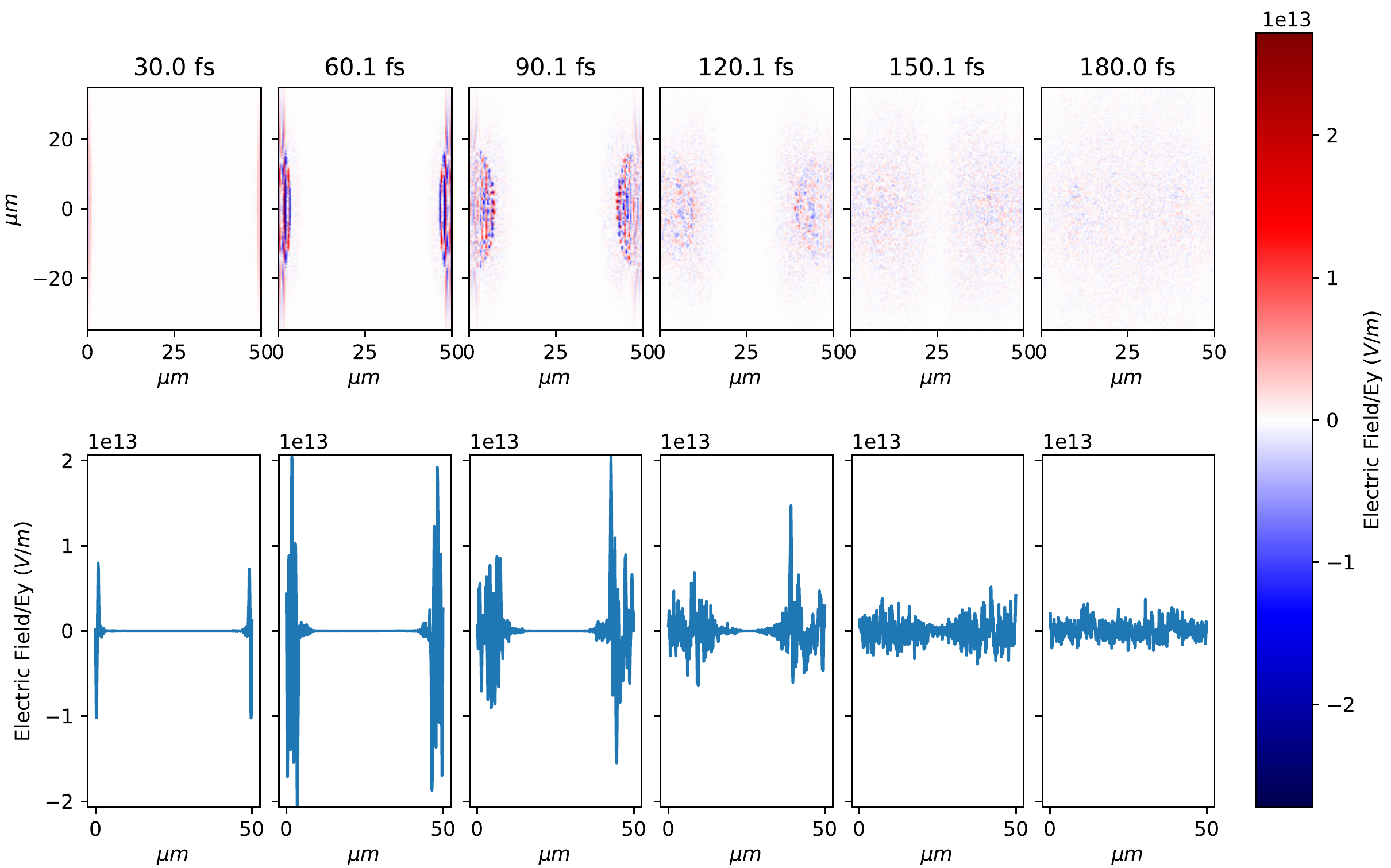}}
\end{center}
\vskip -3mm
\caption{(color online) 
The same as Fig. \ref{A-test-1} with target density 
$n_H = 2.13 \cdot 10^{21} $ atoms$/ $cm$^3$. At this higher density the
energy of EM radiation is absorbed by the ionization of the target. The upper set of figures show the Electric field in the whole target, while the lower set of figures show the E-field in the middle of the target.
}
\label{A-test-2}
\end{figure*}

The EM field strength for higher initial target densities, 
$n_H = 10^{21}$/cm$^3$,
decreases strongly and becomes random. This is the consequence
of the strong absorption by the denser target.  We can see the signs of 
increasing randomness in Fig. \ref{A-test-2}. 
The amplitude of both $E_y$ and $B_z$ is
reduced and random fluctuations increase. This can be 
seen already in Fig. \ref{A-test-2}, where at 150 fs,
the penetration of the $B_z$ field into the target is 
delayed and reduced, and the constructive interference of
the two opposing $B_z$ fields is delayed.

This can be attributed to the kinetic approach, where 
the interactions and pressure in the target are neglected
and thus the dominant longitudinal momentum is transported
to the kinetic motion of the target particles.
These then 
contribute to the pinch effect reducing the EM-field strength
and target beam directed momentum. 

The model includes a kinetic collision set-up, which
reproduces the relaxation time approximation, so that the
momentum distribution converges towards the Maxwell-Boltzmann
ideal thermal distribution. This can then be characterized 
by a temperature.  At the same time this effect is demonstrated 
for collision within the same type of particles.

Thus it is more realistic for our modeling to consider the
target and projectile atoms and ions as separate particle 
species and different type of marker particles (simulation
particles: $H_p,\ H_t,\ p_p,\ {\rm and}\ p_t$, while the electrons
may remain in a single group, as these are more
spread out in the space due to their smaller mass.

The LWF waves show increasing penetration
into the target with increasing laser beam energy. This
is also reconfirmed by recent experiments 
\cite{KBGSF2020}, 
showing that with increasing laser beam energy the reflection
of the beam decreases, while the absorption increases to 100\%
in case of $Au$ target.
 
One should check actually the energy density distribution and its 
time dependence, in this dynamical situation. Up to now in the
target studies 
\cite{CMea2020}
mainly static final state configurations were discussed.

 As the pressure and the Equation of State (EoS) does not appear in
the EPOCH code, it is clearly solving the kinetic equations
of motion for the electrons and ions. Relativistic thermal distribution
functions are included into the code.

In dense nuclear plasma, the EoS is also vital, as at lower densities
we have nuclear attraction while at higher densities strong nuclear 
repulsion. In ideal gas kinetic theory these essential effects are
not taken into account. This could play a significant role in correct
estimate of the burning reaction rate of the target.

An experimental reconfirmation of the effect of energetic laser beams 
was done on a gold target
\cite{KBGSF2020}, 
recently, and strong light absorptions as well as the conversion 
effect to $X$-rays was observed. This experiment
actually confirms earlier, original ideas in
\cite{CS2015,CsKP2018,CsEA2020}.

In \cite{KBGSF2020}, the conversion to $X$-rays radiation is parametrized
by a "velocity", attributing the  frequency shift to a Doppler effect. 
Nevertheless, this velocity is not connected to any of the several 
possible and above mentioned processes, and the authors do not
elaborate what this velocity parameter would characterize.

\section{Outlook for Nanoantennas}
\label{EoN}

The penetration of the LWF waves into the target was mentioned earlier,
and in colliding beam configuration this leads to substantial ion density
increase. The role of nanoantennas in this case is that the increased
photon absorption leads to increased momentum deposition in the target
and higher density
\cite{CsEA2020}.
Finally this results in faster burning rate.

The nanoantennas have another effect too. Similarly to the golden
(or depleted Uranium) hohlraum, which converts the incoming laser light,
to $X$-rays. In case of internal nanoantennas in the target DT fuel
we have two main effect for conversion of the visible laser light to
higher $X$-ray frequencies: Bremsstrahlung in electron collisions, and 
transition from high energy level electron states to lower ones.

This second effect is especially strong for resonant nanoantennas, which
lead to periodic extreme high electron densities at the edge of the antenna.
As the electrons are Fermions, at high density they are forced to occupy
high energy levels (at the cost of the incoming laser energy) and then
as the density periodically decreases they emit the corresponding $X$-rays.
This process is specific to nanoantennas. Due to momentum conservation
the incoming laser light momentum and the emitted $X$-ray momentum are
relatively aligned. 

A recent experiment shows 
\cite{KBGSF2020},
that even without nanoantennas, laser light with increasing intensity 
on gold target leads to increasing absorption (and vanishing reflection).
Furthermore, at the same time the increasing intensity laser irradiation
is accompanied by increasing x-ray conversion. Embedded {\bf resonant}
nanoantennas are expected to amplify these effects.

Resonant nanoantennas have an electron density fluctuation parallel to the 
oscillation of the EM field of the laser irradiation. 
Significant charge separation accompanies the plasmonic resonance, which leads to $10^8$ C/m$^3$ average charge density on the nanorod in linear approximation at the peak of a 26 fs pulse already at $1.4 \cdot10^{12} $W/cm$^2$ that was reported as a threshold resulting in permanent damage of similar gold antennas \cite{BJNagy2020}. Further studies are in progress to determine the charge separation by taking the nonlinear phenomena arising at high intensities into account \cite{CMea2020}.
This indicates that the electron density 
increase due to nanoantennas will 
significantly contribute to the conversion to $X$-rays.

The Bremsstrahlung effect is always present in electron collisions with,
atoms, ions and other electrons. 
Also electron collisions on nano particles can be taken into account as
an additional effect of nanoantannas. More importantly at the point when
all atoms are ionized the nanoantannas do break apart also with dynamical
domains of high electron density. These domains lead to additional collisions
and Bremsstrahlung. Thus the internal resonant nanoantennas have an 
increased conversion of visible light to $X$-rays.


The short and more energetic ignition laser pulse
modifies the thermal equilibrium ignition
scenario. There
the average cross section is calculated as
the thermal average of the constituent fuel
ions (i), i.e. D and T ions. In thermal average
the thermal energy of different components are
equal, 
$E^k_i \approx 3/2 \cdot kT$.  See e.g. 
\cite{Bonasera2019,Barbarino2015}. 
Thus the average 
speed, the relative collision speed  for 
heavier constituents is considerably smaller
(e.g. for protons about 1800 times less than 
for electrons). Consequently the thermal
average collision rate is relatively small
and increases around 
$E^k_{Thermal} \approx 100-200$keV 
\cite{Barbarino2015}.

In contrast for colliding beams with Laser
Wake Field waves, electrons  and ions move
with the same Laser Wake (LW) speed, $v_{LW}$.
So, if the projectile and target interpenetrate 
each other before equilibration, the relative 
energy of ions are larger, (e.g. for 
protons (p) and electrons (e): 
$E^k_p \approx 1800 E^k_e$
). Furthermore, the
relative (r) kinetic energy between projectile 
and target ions is twice as much, 
(e.g. $E^r_p \approx 3600 E^k_e$). 
This is the initial ignition mechanism. After 
3-4 subsequent collision equilibration is
reached, and the relative speed becomes the 
much smaller, isotropically distributed thermal
speed. However, in case of an ignition laser 
pulse of  $\sim$10 fs, there is no time for such
equilibration, until the burning becomes a 
dominant process. For longer, ns, laser pulses
two step nuclear processes are considered 
\cite{Barbarino2015}, 
but for $\sim$10 fs pulses these are negligible.

Thus, it is important to study experimentally, how the LWF waves and the
nanoantenna distribution can achieve the best time-like ignition
in most of the whole target!

\begin{figure*}[!h]  
\begin{center}
\resizebox{0.9\textwidth}{!}
{\includegraphics{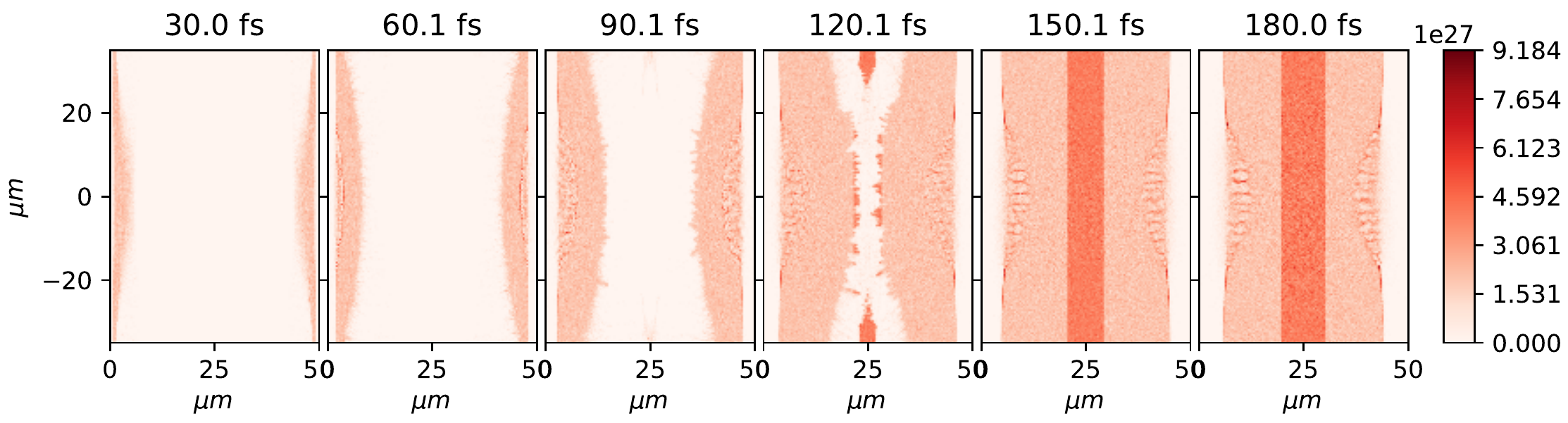}}
\end{center}
\caption{(color online) 
The density of electrons in 1/m$^3$ units. The electron density reaches 
$n_e = 3-5 \cdot 10^{27} $ electrons$/ $m$^3$,
at the initial $H$ atom
target density of $n_H = 2.13 \cdot 10^{21} $ atoms $/ $cm$^3$.
}
\label{A-test-3}
\end{figure*}

Interactions of electrons on the nanoplasmonic surfaces, which 
can interact with the surrounding $H$ ions may lead eventually
to $e(p,n)\nu_e$ reactions in case of high electron density. The
low energy neutrons then can form Deuterium ions, $H^+(n,\gamma)D^+$.
See: https://www-nds.iaea.org/ngatlas2/

\section*{Acknowledgments}
Enlightening discussions with Alex C. Hoffmann and Oliver A. Fekete
are gratefully acknowledged.
Horst St\"ocker acknowledges the Judah M. Eisenberg Professor Laureatus chair
at Fachbereich Physik of Goethe Universit\"at Frankfurt.
D\'enes Moln\'ar acknowledges support by the US Department of Energy, Office of Science, under Award No. DE-SC0016524. We would like to thank the Wigner GPU Laboratory at the Wigner Research Center for Physics for providing support in computational resources.
This work is supported in part by the Institute of Advance Studies, 
K{\H o}szeg, Hungary, 
the Frankfurt Institute for Advanced Studies, Germany,
the E\"otv\"os, Lor\'and Research Network of Hungary, 
the Research Council of Norway, grant no. 255253, and
the National Research, Development and Innovation Office of Hungary,
projects:
Optimized nanoplasmonics (K116362), and
Ultrafast physical processes in atoms, molecules, nanostructures 
and biological systems (EFOP-3.6.2-16-2017-00005). 



\begin{thebibliography}{99}

\bibitem{CsEA2020}
L.P. Csernai, M. Csete, I.N. Mishustin, A. Motornenko, I. Papp, L.M. Satarov, H. St\"ocker \& N. Kro\'o,
Radiation-Dominated Implosion with Flat Target,
{\it Physics and Wave Phenomena}, {\bf 28} (3) 187-199 (2020) in press,
accepted February 3, 2020, (arXiv:1903.10896v3).

\bibitem{CsKP2018}
L.P. Csernai, N. Kro\'o, \& I. Papp,
Radiation-Dominated Implosion with Nano-Plasmonics,
{\it Laser and Particle Beams} {\bf 36}, 171-178 (2018).

\bibitem{Cs1987} 
L.P. Csernai,
Detonation on Timelike Front for Relativistic Systems,
School of Physics, University of Minnesota, Minneapolis, Minnesota, USA,
{\it Zh. Eksp. Teor. Fiz.} {\bf 92}, 397-386 (1987), \& 
{\it Sov. Phys. JETP} {\bf 65}, 219 (1987).

\bibitem{CS2015} 
L.P. Csernai, \& D.D. Strottman,
Volume Ignition via Time-Like Detonation in Pellet Fusion,
{\it Laser and Particle Beams} {\bf 33}, 279-282 (2015).

\bibitem{KR2016}
N. Kro\'o \& P. R\'acz,
Plasmonics - The Interaction of Light with Metal Surface Electrons,
{\it Laser Physics} {\bf 26}, 084011 (2016).

\bibitem{CMea2020}
M. Csete, A. Szenes, E. T\'oth, O. Fekete, D. Vass, B. B\'anhelyi, L. P. Csernai, N. Kro\'o: Plasmonically enhanced target design for inertial confinement fusion, prepared for publication in \textit{Nanomaterials}, 2020.


\bibitem{Faure2006}
J. Faure, C. Rechatin, A. Norlin, A. Lifschitz, Y. Glinec \& V. Malka,
Controlled injection and acceleration of electrons in
plasma wakefields by colliding laser pulses,
{\it Nature Lett.} {\bf 444}, 737 (2006).

\bibitem{Hansson2016}
M. Hansson, B. Aurand, H. Ekerfelt, A. Persson \& O. Lundh,
Injection of electrons by colliding laser pulses in a laser
wakefield accelerator,
{\it Nucl. Instr. and Methods} A {\bf 829}, 99-103 (2016).

\bibitem{Bar-Lev2014}
Doron Bar-Lev and Jacob Scheuer,
Plasmonic metasurface for efficient ultrashort pulse 
laser-driven particle acceleration,
{\it Phys. Rev. STAB} {\bf 17}, 121302 (2014).

\bibitem{Bonasera2019}
G. Zhang, M. Huan, A. Bonasera, Y. G. Ma, B. F. Shen, H. W. Wang, J. C. Xu, G. T. Fan, H. J. Fu, H. Xue, H. Zheng, L. X. Liu, S. Zhang, W. J. Li, X. G. Cao, X. G. Deng, X. Y. Li, Y. C. Liu, Y. Yu, Y. Zhang, C. B. Fu, and X. P. Zhang,
Nuclear probes of an out-of-equilibrium plasma at the highest compression,
{\it Phys. Lett.} A {\bf 383} (19), 2285-2289 (2019). 

\bibitem{Jung2015}
D. Jung, B.J. Albright, L. Yin, D.C. Gautier, B. Dromey, R. Shah,
S. Palaniyappan, S. Letzring, H.-C. Wu, T. Shimada, R.P. Johnson, D. Habs,
M. Roth, J.C. Fernandez, and B.M. Hegelich,
Scaling of ion energies in the relativistic-induced
transparency regime,
{\it Laser and Particle Beams} {\bf 33}, 695--703 (2015).

\bibitem{Arber2015}
T. D. Arber, et. al.Contemporary particle-in-cell approach to laser-plasma modelling {\it Plasma Phys. Control. Fusion} {\bf}, 57, 113001 (2015)

\bibitem{FLZhang2011}
F.L. Zheng, S.Z. Wu, C.T. Zhou, H.Y. Wang, X.Q. Yan and X.T. He,
An ultra-short and TeV quasi-monoenergetic ion
beam generation by laser wakefield accelerator in
the snowplow regime,
{\it Europhys. Lett. (EPL)}, {\bf 95}, 55005 (2011).

\bibitem{KBGSF2020}
Zs. Kov\'acs, K. Bali, B. Gilicze, S. Szatm\'ari and I.B. F\"oldes,
Reflectivity and spectral shift from laser plasmas generated
by high-contrast, high-intensity KrF laser pulses,
{\it Phil. Trans. R. Soc.} A {\bf 378}, 20200043 (2020).


\bibitem{BJNagy2020}
B. J. Nagy et al.: Near-Field-Induced Femtosecond Breakdown of Plasmonic Nanoparticles, \textit{Plasmonics} \textbf{15}, 335--340 (2020)

\bibitem{Barbarino2015}
M. Barbarino, Fusion reactions in laser produced plasma, PhD thesis 
Texas A\&M University (2015).


\end{thebibliography}
\end{document}